\documentclass{ws-p8-50x6-00}
\usepackage{amsmath}
\usepackage{amscd}
\usepackage{amssymb}
\usepackage{array}

\begin{document}

\newcommand{\C}{\mathbb{C}}
\newcommand{\R}{\mathbb{R}}
\newcommand{\Ha}{\mathbb{H}}
\newcommand{\A}{\frak{a}}

\title{Spinor Algebras and Extended Superconformal Algebras}

\author{M. A.  Lled\'o}

\address{Dipartimento di Fisica, Politecnico di
Torino, Corso Duca degli Abruzzi 24, I-10129 Torino, Italy, and
INFN, Sezione di Torino, Italy.\\ E-mail: lledo@athena.polito.it}

\author{V. S. Varadarajan}

\address{Department of Mathematics, University of
California, Los Angeles, Los Angeles, CA 90095-1555, USA.\\
E-mail: vsv@math.ucla.edu}

\maketitle

\abstracts{ We consider supersymmetry algebras in arbitrary
spacetime dimension and signature. Minimal and maximal
superalgebras are given for single and extended supersymmetry. It
is seen that the supersymmetric extensions are uniquely determined
by the properties of the spinor representation, which depend on
the dimension $D$ mod 8 and the signature $|\rho|$ mod 8 of
spacetime. }

\section{ Spacetime Symmetry Algebras and the Super Lie Algebras Containing Them}

The idea that one can have a new type of algebraic structure
generating symmetries between bosonic and fermionic states goes
back to the 1970's when several physicists discovered what are
nowadays known as {\it super Lie algebras.\/} The super Lie
algebra contains the usual spacetime symmetry algebra (Poincar\'e
or conformal algebra) and its representations combine several
representations of the spacetime algebra into a single module
irreducible under the super algebra, thus giving rise to the
concept of a supermultiplet. The central problem is therefore the
following: given  a spacetime symmetry algebra, to obtain the
super Lie algebras that contain the given Lie algebra in its even
part.

By spacetime we mean  a real vector space with a nondegenerate
quadratic form of type $(s,t)$. It is denoted by $V(s,t)$. The
dimension of spacetime is denoted by $D=s+t$, and the signature of
the quadratic form by $\rho=s-t$. The group of coordinate
transformations in $V$ that leaves the metric invariant up to
scale transformations is the simple group SO($s+1,t+1$) (for
$D=s+t\geq 3$), being the Poincar\'e group ISO$(s,t)$ a subgroup
of it. The Poincar\'e group ISO$(s,t)$ can also be obtained as a
contraction of the conformal group in one dimension less SO($s,
t+1$).

As an example, we can take a spacetime of type $(s,t)=(10,1)$. We
have $$\begin{CD}\frak{so} (10,2)@>{\rm contraction }>>\frak{
iso}($10,1$)@>{\rm subalgebra}>>\frak{ so}(11,2).
\end{CD}$$
The supersymmetric extension of the conformal algebra in $D=10$ is
$\frak{osp}(1|32)$, which gives as a contraction the M-theory
superalgebra, that is, the super Poincar\'e algebra with five and
two-brane charges. This can be seen as a subalgebra of
$\frak{osp}(1|64)$, the superconformal algebra in $D=11$.

For a fixed dimension $D$, theories defined on spacetimes with the
same signature $|\rho|$ mod 8 may be related by some duality since
the superalgebras are the same. As an example, the theories
introduced by Hull \cite{hu}, M, M$^*$ and M', are formulated in
11 dimensional spacetimes with signatures $(s,t)=(10,1),
(9,2),(6,5)$ respectively. In all these theories the signature of
spacetime is $\rho=s-t=\pm1$ mod 8.

In this paper we will see how the supersymmetric extensions of
spacetime superalgebras are determined by $D$ and $|\rho|$ mod 8.
All these  results are contained in  \cite{dflv,dfl}.

\section{  Spin Groups and Spin Modules.}

For each spacetime $V(s,t)$ there  are two spin groups,
Spin$(s,t)$ and Spin$_{\C}(s,t)$, the real and complex spin
groups, the former being a real form of the latter. As it is well
known, when $D$ is odd there is only one irreducible spin module
of  the complex group. It will be denoted by $S$. When $D$ is even
there are two different irreducible spin modules denoted by $S^\pm
$ (chiral spin modules). For constructing superspacetimes and
superfields we need real forms of the complex spin modules. These
do not always exist and sometimes we have to combine several
copies of them to get a real irreducible module for Spin$(s,t)$.

The reality issues depend only on $|\rho| $ mod 8. The type of a
real irreducible module is the isomorphism class of the commutant
of the module. Since this commutant is a division algebra, it has
to be one of ${\R}, {\C}, {\Ha}$. For $\rho \equiv 0,1,7$ the
irreducible real spin modules are actually real forms of the
complex irreducible spin modules. For $\rho \equiv 3,4,5$ they are
quaternionic and on complexification become $2S^\pm $ or $2S$ (in
the even and odd $D$ cases respectively). For $\rho \equiv 2,6$
they are of complex type and their complexifications are
$S^+\oplus S^-$.

 The spin
module may admit an invariant form. If this form is symmetric (it
is unique up to multiplication by a scalar), then the spin module
is said to be orthogonal. If it is antisymmetric the spin module
is symplectic and if  it  has no invariant form it is called
linear. The complex spin groups are embedded into the complex
orthogonal, symplectic and linear groups respectively. The
existence and symmetry properties of this form depend only on $D$
mod 8. These mod 8 results have been known for a long time both to
physicists and mathematicians, but there is now a unified approach
based on the super Brauer group \cite{wa,de}.

\section{Non extended supersymmetry}

\subsection{Super Poincar\'e Algebras}

A non extended ($N=1$) super Poincar\'e algebra is a super Lie
algebra whose even part is the Poincar\'e algebra, the odd part
consists of one real irreducible spin representation  and the
anticommutator of two spinors (elements of the spin module) is
proportional to the momentum. If we denote a spinor in the form
$Q_\alpha$  we have that the anticommutator $\{Q_\alpha,Q_\beta\}$
is symmetric in the indices $\alpha, \beta$. The biggest Lie
algebra that one can put in the right hand side of the
anticommutator is the Lie algebra of symmetric matrices,
$\frak{sp}(2n)$ ($2n$ is the real dimension of the spin module).
In fact, there is a superalgebra which has $\frak{sp}(2n)$ as
bosonic part, it is   the orthosymplectic algebra
$\frak{osp}(1|2n)$.

In order to see if the required extension exists, we have to
determine if there is an  abelian subalgebra of $\frak{sp}(2n)$
with the commutation rules of a vector with the generators of  the
orthogonal algebra. In other words, we have to determine if there
is a morphism of Spin$(s,t)$ modules $$ S\otimes S\longrightarrow
V.$$ (It is easy to see that this condition is also sufficient).

In fact one can study the question for existence of maps $S\otimes
S\longrightarrow \Lambda ^kV$ for all $k$. As we will see in the
next subsection, the case $k=2$ is important for constructing
super conformal algebras, namely super Lie algebras extending
$\frak s\frak o (s+1,t+1)$.

The maps $S\otimes S\longrightarrow \Lambda ^kV$ are unique
(projectively) if they exist and they are either symmetric (+) or
antisymmetric (-). The full details can be worked out without
difficulty using the formalism of Deligne \cite{de} and are listed
in Table \ref{morphisms}.

\begin{table}[ht]
\begin{center}
\begin{tabular} {|c |c|c||c|c|}
\hline \multicolumn{1}{|c|}{$D$} &\multicolumn{2}{| c||}{$k $
even}&\multicolumn{2}{|c|}{$k $ odd}\\\hline\hline & morphism &
symmetry &morphism&symmetry\\
  \hline
  0& $S^\pm\otimes S^\pm\rightarrow\Lambda^k$&$(-1)^{k(k-1)/2}$& $S^\pm\otimes
  S^\mp\rightarrow\Lambda^k$&\\\hline
  1& $S\otimes S\rightarrow\Lambda^k$&$(-1)^{k(k-1)/2}$& $S\otimes
  S\rightarrow\Lambda^k$&$(-1)^{k(k-1)/2}$\\\hline
  2& $S^\pm\otimes S^\mp\rightarrow\Lambda^k$&& $S^\pm\otimes
  S^\pm\rightarrow\Lambda^k$&$(-1)^{k(k-1)/2}$\\\hline
  3& $S\otimes S\rightarrow\Lambda^k$&$-(-1)^{k(k-1)/2}$& $S\otimes
  S\rightarrow\Lambda^k$&$(-1)^{k(k-1)/2}$\\\hline
  4& $S^\pm\otimes S^\pm\rightarrow\Lambda^k$&$-(-1)^{k(k-1)/2}$& $S^\pm\otimes
  S^\mp\rightarrow\Lambda^k$&\\\hline
  5& $S\otimes S\rightarrow\Lambda^k$&$-(-1)^{k(k-1)/2}$& $S\otimes
  S\rightarrow\Lambda^k$&$-(-1)^{k(k-1)/2}$\\\hline
  6& $S^\pm\otimes S^\mp\rightarrow\Lambda^k$&& $S^\pm\otimes
  S^\pm\rightarrow\Lambda^k$&$-(-1)^{k(k-1)/2}$\\\hline
  7& $S\otimes S\rightarrow\Lambda^k$&$(-1)^{k(k-1)/2}$& $S\otimes
  S\rightarrow\Lambda^k$&$-(-1)^{k(k-1)/2}$\\\hline

\end{tabular}
\caption{Properties of morphisms.}\label{morphisms}
\end{center}
\end{table}

We denote $\rho=\rho_0+n8$, $D=D_0+m8$, with $0\leq \rho_0,D_0<7$,
$m$ and $n$ integers.

\medskip

\noindent{\it Real Case}. There is a conjugation $\sigma$ on $S$
commuting with the action of the spin group, so there is a real
form of  $S$ which is a real Spin$(s,t)$-module. The
anticommutator of two spinors is
$$\{Q_\alpha,Q_\beta\}=\sum\limits_{k} \gamma^{[\mu_1\cdots
\mu_k]}_{(\alpha \beta)}Z_{[\mu_1\cdots \mu_k]},$$ so the morphism
$\gamma^{\mu}_{\alpha\beta}$ must be symmetric in
 $\alpha, \beta$.
This happens for $$
\begin{aligned}
&\rho_0=0,\\
 &\rho_0=1,7,\end{aligned}\qquad
 \begin{aligned}
&D_0=2\\ &D_0=1,3.\end{aligned}$$
 If we consider also non chiral superalgebras (with
$S^\pm$ both present), the we find that for $\rho_0=0$ and
$D_0=0,4$ there is a morphism $S^+\otimes S^-\rightarrow V$, and
the extension is possible

\medskip

\noindent{\it Quaternionic case}. There is a pseudoconjugation
$\sigma$ on $S$ commuting with Spin$(s,t)$. We need two copies
$S\oplus S=S\otimes\C^2$ to be able to construct a conjugation
$\tilde\sigma=\sigma\otimes\sigma_0$ and then to impose a reality
condition. The factor
 $\C^2$ is an internal index space. The
biggest group that commutes with $\sigma_0$ is $\rm{SU}(2)\simeq
\rm{USp}(2)\simeq \rm{SU}^*(2)$.

In the anticommutator the internal index appears,
$$\{Q^i_\alpha,Q^j_\beta\}=\sum\limits_{k} \gamma^{[\mu_1\cdots
\mu_k]}_{[\alpha \beta]}Z^0_{[\mu_1\cdots \mu_k]}\Omega^{ij}+
\sum\limits_{k} \gamma^{[\mu_1\cdots \mu_k]}_{(\alpha
\beta)}Z^I_{[\mu_1\cdots \mu_k]}\sigma_I^{ij},$$ where $\Omega$ is
the symplectic matrix and $\sigma_I$ are the symmetric Pauli
matrices, $I=1,2,3$.

If we impose invariance under SU(2) (R-symmetry group) only the
first term (SU(2) singlet) may appear and the morphism
$\gamma^{\mu}_{\alpha \beta}$ must be antisymmetric. This happens
for $$\begin{aligned} &\rho_0=3,5,\\
&\rho_0=4\\&\rho_0=4\end{aligned}\qquad\begin{aligned}
&D_0=5,7\\&D_0=6\\&D_0=0,4 \quad \hbox{(non
chiral).}\end{aligned}$$ If we want to take the generators in the
second term, then the $R$-symmetry is broken. For example, taking
$\delta^{ij}$ breaks the symmetry to the subgroup SO$^*$(2). Then
we have superalgebras for $$\begin{aligned}
&\rho_0=3,5,\\&\rho_0=4\\&\rho_0=4&\end{aligned}\qquad
\begin{aligned}&D_0=1,3\\&D_0=2\\&D_0=0,4\quad \hbox{(non
chiral)}.\end{aligned}$$

Both types of R-symmetry will find an interpretation later. We
note that non chiral algebras appear with both, SU(2) symmetry and
SO$^*$(2) for the same space times. In fact, one can easily see
that SO$^*$(2)$\subset$SU(2) and that they are isomorphic super
algebras.

\medskip

\noindent{\it Complex case}. There is no conjugation nor
pseudoconjugation. We take the direct sum $S^+\oplus S^-$, and
there is a conjugation with the property $\sigma(S^\pm)=S^\mp$.
Again we have an $R$-symmetry group, U(1), with $S^\pm$ having
charges $\pm 1$. If we want invariance under U(1) the orthogonal
generators must appear in the anticommutator $\{Q^+,Q^-\}$, so the
morphism should be of type $$S^\pm\otimes S^\mp\longrightarrow
V.$$ This happens for $\rho_0=2,6, \; D_0=0,4$.

 Otherwise,
the morphisms can be of type $S^\pm\otimes S^\pm\rightarrow
\Lambda^2,$  (symmetric) which happens for $\rho_0=2,6,\; D_0=2$.

\medskip

We note that for physical spacetimes there is always the
possibility of supersymmetric extension.

\section{Super Conformal Algebras}

A super conformal algebra is a real, simple superalgebra whose
even part contains the conformal group. The odd part of the super
conformal algebra is a direct sum of $N$ real spinor modules.

A simple superalgebra $\A=\A_0\oplus \A_1$ has the propriety
$$\{\A_1,\A_1\}=\A_0.$$ For a fixed $N$, depending on how big is
$\A_0$ we can have from minimal to maximal superconformal
algebras.

As in the case of  Poincar\'e, a necessary condition is that
\begin{equation}\frak{o}(s,t)\subset\frak{sp}(2n).\label{inclusion}\end{equation}
 The adjoint of
$\frak{o}(s,t)$ is  the two fold antisymmetric representation, so
one should look for the symmetries of morphisms $$S\otimes
S\longrightarrow \Lambda^2.$$ $\frak{osp}(1|2n)$ is a simple super
algebra; it is in fact the
  maximal super conformal algebra in the cases where condition
  (\ref{inclusion}) is  satisfied.  This happens for
 $$ \begin{aligned}& \hbox{Real case:}\\
 &\\
 & \hbox{Quaternionic case:}\\&
 \\&\\&\\&\hbox{Complex case:}\\&\end{aligned}\quad
\begin{aligned}&\rho_0=0\\&\rho_0=1,7\\&\rho_0=0\\&\rho_0=4\\&\rho_0=3,5,\\&\rho_0=4
\\&\rho_0=4,\\&\rho_0=3,5,\\&\rho_0=2,6
\\&\rho_0=2,6\end{aligned}\quad
 \begin{aligned}&D_0=4\\&D_0=3,5\\&D_0=2,6 \quad\hbox{(non chiral)}\\&D_0=8\\&D_0=1,7\\&D_0=2,6\quad\hbox{(non chiral)}\\&D_0=4
 \quad {(\rm SO^*(2))}\\&D_0=3,5\quad {(\rm SO^*(2))}\\&D_0=2,6\\&D_0=4\quad\hbox{(no
 U(1))}.
\end{aligned}$$

As we have mentioned, the complex  group is embedded into a larger
group, depending whether the spinor representation is orthogonal,
symplectic or linear. The spin representation goes into the
fundamental representation of the corresponding group (orthogonal,
symplectic or linear). The morphisms $$S\otimes S\longrightarrow
\Lambda^0,$$ being the invariant bilinear forms, their existence
is governed by $D$ (mod 8).

The interesting question is to see if there is a real form of
these groups containing the real Spin$(s,t)$ group. We have to
play with the reality properties of the spinors, which depend on
$\rho$, so we have all the possibilities for $D$ and $\rho$. The
result is that the real form is uniquely determined. It is listed
in Table \ref{min}.

To make an example, let $\rho_0=1,7$ and $D_0=1,7$. $\dim
(S)=2^{D-1/2}$ and one can show that $$ {\rm
Spin}(V)\longrightarrow {\rm SO}(2^{(D-3)/2}, 2^{(D-3)/2}) \qquad
(\min (s,t)\ge 1). $$ If $D_0=3,5$, the spin group goes inside a
symplectic group and in fact we have $$ {\rm Spin}
(V)\longrightarrow {\rm Sp}(2^{(D-1)/2}, {\R}).$$

Some low dimensional examples are
\begin{eqnarray*}{\rm Spin}(4,3)\subset {\rm SO}(4,4), \qquad {\rm
Spin}(8,1)\subset {\rm SO}(8,8)\\ {\rm Spin}(2,1)\subset {\rm
SL}(2,\R), \qquad {\rm Spin}(3,2)\subset {\rm Sp}(4,\R)\\ {\rm
Spin}(3)\subset {\rm SU}(2), \qquad {\rm Spin}(5)\subset {\rm
USP}(4), \\ {\rm Spin}(4,1)\subset {\rm USp}(2,2) \qquad {\rm
Spin}(4,3)\subset {\rm SO}(4,4), \\ {\rm Spin}(8,1)\subset {\rm
SO}(8,8).\end{eqnarray*}

The Lie algebra of this group is called the Spin$(s,t)$-algebra.
It happens that for each combination of $\rho$ and $D$ from the
ones listed above there is a simple superalgebra whose even part
is $${\rm Spin}(s,t)-{\rm algebra}\otimes R-{\rm symmetry}.$$ This
superalgebra is minimal inside the classical series, and the list
of superalgebras is given in Table \ref{min}. There is an
exceptional superalgebra that is smaller for dimension 5, with
bosonic part $\frak{so}(5,2)\oplus\frak{su}(2)$, it is the
exceptional super algebra $\frak{f}(4)$. For spacetimes of
dimensions 3, 4 and 6 we also have the conformal algebra as a
factor in the bosonic part of the superalgebra. We have
\begin{gather*}\frak{sp}(4,\R)\approx\frak{so}(3,2)\\
\frak{su}(2,2)\approx\frak{so}(4,2)\\
\frak{so}^*(4,\R)\approx\frak{so}(6,2).
\end{gather*}

\begin{table}[ht]
\begin{center}
\begin{tabular} {|c|c|l|l|}
\hline
 $D_0$   & $\rho_0$& Spin($V$) algebra&Spin($V$) superalgebra\\\hline\hline
 1,7& 1,7& $\frak{so}(2^{(D-3)/2},2^{(D-3)/2})$&  \\\hline
 1,7& 3,5 & $\frak{so}^*$($2^{(D-1)/2})$&$\frak{osp}(2^{(D-1)/2})^*|2)$ \\\hline
 3,5& 1,7& $\frak{sp}(2^{(D-1)/2},\R$)&$\frak{osp}(1|2^{(D-1)/2},\R)$\\\hline
 3,5& 3,5&
 $\frak{usp}(2^{(D-3)/2},2^{(D-3)/2}$)&$\frak{osp}(2^*|2^{(D-3)/2},2^{(D-3)/2}$)
  \\\hline\hline
 0& 0&$\frak{so}(2^{(D-4)/2},2^{(D-4)/2}$) &  \\\hline
 0& 2,6&$\frak{so}(2^{(D-2)/2},\C)^\R$ &  \\\hline
 0& 4 &$\frak{so}^*(2^{(D-2)/2}$)&$\frak{osp}(2^{(D-2)/2})^*|2)$\\\hline
 2,6&  0&$\frak{sl}(2^{(D-2)/2},\R)$&$\frak{sl}(2^{(D-2)/2}|1, \R)$ \\\hline
 2,6&2,6& $\frak{su}(2^{(D-4)/2},2^{(D-4)/2})$&$\frak{su}(2^{(D-4)/2},2^{(D-4)/2}|1)$\\\hline
2,6 & 4 &$\frak{su}^*(2^{(D-2)/2}))$&$\frak{su}(2^{(D-2)/2})^*|2)$
\\\hline
  4&0&$\frak{sp}(2^{(D-2)/2},\R$)&$\frak{osp}(1|2^{(D-2)/2},\R)$ \\\hline
 4&2,6&$\frak{sp}(2^{(D-2)/2},\C)^\R$&$\frak{osp}(1|2^{(D-2)/2},\C)$ \\\hline
 4&4&$\frak{usp}(2^{(D-4)/2},2^{(D-4)/2}$)&  $\frak{osp}(2^*|2^{(D-4)/2},2^{(D-4)/2}$)\\\hline
\end{tabular}
\caption{Minimal Spin($V$) superalgebras.}\label{min}
\end{center}
\end{table}

 \section{Extended supersymmetry.}

\begin{table}[ht]
\begin{center}
\begin{tabular} {c|c|c|l |l|}

\cline{2-5} &$D_0$&$\rho_0$& R-symmetry&Spin$(s,t)$
superalgebra\\\cline{2-5} $\circ$& 1,7& 1,7& $\frak{
sp}(2N,\R)$&$\frak{
osp}(2^{\frac{D-3}{2}},2^{\frac{D-3}{2}}|2N,\R)$\\\cline{2-5}
$\star$& 1,7& 3,5& $\frak{usp}(2N-2q,2q)$&$\frak{
osp}(2^{\frac{D-1}{2}\,*}|2N-2q,2q)$\\\cline{2-5} $\star$& 3,5&
1,7& $\frak{so}(N-q,q)$&$\frak{
osp}(N-q,q|2^{\frac{D-1}{2}})$\\\cline{2-5} & 3,5& 3,5& $\frak{
so}^*(2N)$&$\frak{
osp}(2{N}^*|2^{\frac{D-3}{2}},2^{\frac{D-3}{2}})$\\\cline{2-5}
\cline{2-5} $\circ$& 0& 0& $\frak{sp}(2N,\R)$&$\frak{
osp}(2^{\frac{D-4}{2}},2^{\frac{D-4}{2}}|2N)$\\\cline{2-5}
$\circ$& 0& 2,6& $\frak{sp}(2N,\C)_\R$&$\frak{
osp}(2^{\frac{D-2}{2}}|2N,\C)_\R$\\\cline{2-5} $\star$& 0& 4&
$\frak{usp}(2N-2q,2q)$&$\frak{
osp}(2^{\frac{D-2}{2}\,*}|2N-2q,2q)$\\\cline{2-5} & 2,6& 0&
$\frak{ sl}(N,\R)$&$\frak{
sl}(2^{\frac{D-2}{2}}|N,\R)$\\\cline{2-5} $\star$&2,6& 2,6&
$\frak{su}(N-q,q)$&$\frak{
su}(2^{\frac{D-4}{2}},2^{\frac{D-4}{2}}|N-q,q)$\\\cline{2-5} &
2,6& 4& $\frak{su}^*(2N,\R)$&$\frak{
su}(2^{\frac{D-2}{2}}|2N)^*$\\\cline{2-5} $\star$& 4& 0& $\frak{
so}(N-q,q)$&$\frak{osp}(N-q,q|2^{\frac{D-2}{2}})$\\\cline{2-5} &
4& 2,6& $\frak{so}(N,\C)_\R$&$\frak{
osp}(N|2^{\frac{D-2}{2}},\C)_\R$\\\cline{2-5} & 4& 4& $\frak{
so}^*(2N)$&$\frak
{osp}(2{N}^*|2^{\frac{D-4}{2}},2^{\frac{D-4}{2}})$\\\cline{2-5}
\end{tabular}
\caption{Spin$(s,t)$ superalgebras.}\label{next}
\end{center}
\end{table}

We want to construct now a super Lie algebra whose odd part
consists of $N$ copies of the spin module. There is again a
maximal super Lie
 algebra, osp$(1/N2n)$, but the presence of the  internal
index space $i=1,\dots N$ allows extra possibilities in the
combinations of $\rho$ and $D$. For example, in the real case we
can have

$$\{Q_\alpha^i,Q_\beta^j\}=\delta^{ij}\sum\limits_{k}
\gamma^{[\mu_1\cdots \mu_k]}_{(\alpha \beta)}Z_{[\mu_1\cdots
\mu_k]},$$ and an orthogonal $R$-symmetry group, or
$$\{Q_\alpha^i,Q_\beta^j\}=\Omega^{ij}\sum\limits_{k}
\gamma^{[\mu_1\cdots \mu_k]}_{(\alpha \beta)}Z_{[\mu_1\cdots
\mu_k]},$$ and  symplectic $R$-symmetry group.

These new possibilities fill in the gaps in Table \ref{min}. The
minimal super Lie algebras have an even part that is again of the
form $${\rm Spin}(s,t)-{\rm algebra}\otimes R-{\rm symmetry}.$$ A
conjugation must exist in the space $S\otimes W$ ($R$ is a module
for the R-symmetry group). If there is a conjugation in $S$, then
there is a conjugation in $W$and  a pseudoconjugation in $S$
implies a pseudoconjugation in $W$. If the spacetime group is
unitary, then the R-symmetry group is also unitary and if the
spacetime group is complex, then the R-symmetry is also complex.

The list is given in Table \ref{next}. We mark with a symbol
``$\star$" the cases that allow the possibility of a compact
R-symmetry group. They correspond to ${\frak so}(s,2)$, that is,
the physical conformal groups. We mark with the symbol ``$\circ$"
the cases that do not arise in the non extended case.

\medskip
\noindent {\it Non chiral superalgebras}. In Tables
\ref{min},\ref{next} we have written a chiral algebra whenever it
is possible and a non chiral one for the rest. Nevertheless, one
can ask if there is a non chiral algebra in the cases where the
chiral one already exists. As an example, consider a spacetime of
type $(5,1)$, that is, $D=6$ and physical signature. The conformal
group is SO(6,2), the orthogonal group of a space time of
dimension 8 and signature 4. There exists a chiral super algebra,
namely $\frak{osp}(8^*|2)$. A non chiral superalgebra can be
constructed if we go to one dimension more and then we make a
dimensional reduction. We can consider the orthogonal groups
SO(7,2) or SO(6,3). In both cases, the associated super conformal
 algebra  is $\frak{osp}(16^*|2)$.
 We have the decomposition
$$\begin{CD}\frak{so}^*(16)@<\supset<<\frak{so}(7,2)@<\supset<<\frak{
so}(6,2)\\
 {\bf 16}@>>>{\bf 16}@>>>{\bf 8}_L +{\bf 8}_R.
\end{CD}$$
So the super conformal algebra in dimension 7 can be seen as a non
chiral superconformal algebra in dimension 6.

\section*{Acknowledgments}
We want to thank our collaborators, S. Ferrara and R. D'Auria
 with whom this work was done.

\end{document}